\documentclass[12pt]{article}
\usepackage{latexsym}
\usepackage{amsmath}
\usepackage{amsfonts}
\usepackage{amssymb}

\newcommand{\n}{{\frac{N+1}{2}}}

\title{N-Galilean conformal algebras and quantum theory with higher order  time derivatives}
\author{K. Andrzejewski\thanks{e-mail: k-andrzejewski@uni.lodz.pl},
J. Gonera, P. Kosi\'nski\\
\small Department of Theoretical Physics and Computer Science, \\
\small University of \L\'od\'z,\\
\small Pomorska 149/153, 90-236 {\L}\'od\'z, Poland
}
\date{}
\begin{document}
\maketitle 
\begin{abstract}
It is shown that centrally extended  N-Galilean conformal algebra, with N-odd, is the maximal symmetry algebra of the Schr\"odinger equation  corresponding to the free Lagrangian involving $\frac{N+1}{2}$-th  order time derivatives.
\end{abstract}
\par
It is well known that Schr\"odinger group is the  maximal symmetry group of free classical motion, while its  central extension is the maximal symmetry group of  the
Schr\"odinger equation of free particle \cite{b10}.
Recently, Gomis and Kamimura \cite{b6} have showed that the free higher-derivative theory 
\begin{equation}
\label{ee1}
\frac{d^{2n}\vec q}{dt^{2n}}=0.
\end{equation}
  defined by the Lagrangian 
\begin{equation}
\label{ee2}
L=\frac{m}{2}\left (\frac{d^n\vec q}{dt^n}\right)^2,
\end{equation}
where $\vec q$ is the  coordinate in d-dimensional  Euclidean space and $m$  is a "mass" parameter of dimension $\textrm{kg}\cdot\textrm{s}^{2(n-1)}$
has a symmetry described by $N$-Galilean conformal algebra  ($N$-GCA)  with $N=2n-1$ (for more information about $N$-GCA and its relations with higher order time derivatives, see \cite{b3}-\cite{b19}  and  references therein). Moreover, they showed that its quantum counterpart,  that  is the Schr\"odinger equation 
\begin{subequations}
\label{ee3}
\begin{equation}
\label{ee3a}
i\partial_t\psi=H\psi,\quad \psi=\psi(t,\vec q^1,\ldots,\vec{q}^{\n})
\end{equation}
where 
\begin{equation}
\label{ee3b}
H=\sum_{j=1}^{\frac{N-1}{2}}\vec p_j\vec q^{j+1}+\frac{1}{2m}\vec p_{\n}^2
\end{equation}
\end{subequations}{
is the  Ostrogradski Hamiltonian \cite{b21} of (\ref{ee2}), exhibits centrally extended  $N$-GCA symmetry. In the previous paper \cite{b0} the authors showed that $N$-GCA is the maximal symmetry algebra of the Lagrangian (\ref{ee2}). Here, we generalize  Niederer's work \cite{b10} and  show that the  centrally extended $N$-GCA is the maximal symmetry algebra of the Schr\"odinger  equation (\ref{ee3}). In order to do this let us recall  that the Lie algebra of the maximal Lie group which does not change equation (\ref{ee3}) under the  change of $\psi$ 
 \begin{equation}
 \begin{split}
 &\psi(t,\vec q^1,\ldots,\vec{q}^{\n})\rightarrow (T_g\psi)(t,\vec q^1,\ldots,\vec q^{\n})=\\
&=f_g(g^{-1}(t,\vec q^1,\ldots,\vec q^{\n}))\psi(g^{-1}(t,\vec q^1,\ldots,\vec q^{\n}))
 \end{split}
 \end{equation}
consist of the operators $X$ 
\begin{equation}
\label{ee5}
-iX(t,\vec q^1,\ldots,\vec q^{\n})=\sum_{j=1}^{\n}\vec a^j\vec\partial_j+a\partial_t+c
\end{equation}
satisfying the following equation
\begin{equation}
\label{ee6}
[P,X]=i\lambda P,   \quad P=i\partial_t-H
\end{equation}
for a certain  function $\lambda=\lambda(t,\vec q^1,\ldots,\vec q^{\n})$.
Let us introduce the following notation 
\begin{equation}
\label{ee7}
\vec q^j=(q_\alpha^j),\quad \vec a^j=(a^j_\alpha), \quad\vec \partial_j=(\partial _j^\alpha),\quad  etc. 
\end{equation}
where $\alpha=1,\ldots,d$  and  (if the contrary is not stated explicitly) repeated indices $i,j,k$ etc. ($\alpha,\beta,\gamma$ etc.) denote summation from  $1$ to $\frac{N-1}{2}$ (from $1$ to $d$, respectively).
\par
In the case of the Hamiltonian (\ref{ee3b}) condition  (\ref{ee6})  implies the following set of equations for coefficients of the operator $X$
\begin{subequations}
\begin{align}
\label{es1}
\lambda&=\partial_t a+q_\alpha^{k+1}\partial_k^\alpha a\\
\label{es2}
\delta^{\alpha}_{\beta}&=\partial^{\alpha}_{\n}a_{\beta}^{\n}+\partial^{\beta}_{\n}a^{\n}_\alpha\\
\label{es3}
0&=i\partial_tc+iq_\alpha^{k+1}\partial_k^\alpha c+\frac{1}{2m}(\partial^\alpha_{\n}\partial^\alpha_{\n})c\\
\label{es4}
0&=i\partial_ta_\alpha^\n+iq_\beta^{k+1}\partial_k^\beta a_\alpha^\n+\frac{1}{2m}(\partial^\beta_{\n}\partial^\beta_{\n})a_\alpha^\n+\frac{1}{m}\partial^\alpha_\n c\\
\label{es5}
0&=\partial_\n^\alpha a\\
\label{es6}
0&=\partial_\n^\alpha a_\beta^j, \quad j=1,\ldots,\frac{N-1}{2}\\
\label{es7}
\lambda q^{j+1}_\beta&=\partial_ta_\beta^j+q_\alpha^{k+1}\partial_k^\alpha a_\beta^j-a_\beta^{j+1}, \quad j=1,\ldots,\frac{N-1}{2}
\end{align}
\end{subequations}
Our main task is to show that the general solution of the above set of equations gives centrally extended N-GCA. In order to simplify our considerations we assume that $N>3$. The case  $N=3$ is simpler and can be obtained in the same way. 
\par
First, let us note that (\ref{es1})  and (\ref{es5}) imply 
\begin{equation}
\label{ee8}
\partial_\n^\beta\partial_\n^\alpha\lambda=0
\end{equation}
On the other hand differentiating (\ref{es7}) (for $j=\frac{N-1}{2}$) with respect to $\partial_\n^\alpha$  (without summation)  and using (\ref{es6}) and (\ref{es2}) 
we obtain 
\begin{equation}
\label{ee9}
\partial_\frac{N-1}{2}^\alpha a_\alpha^\frac{N-1}{2}=\frac 32\lambda+q_\alpha^\n\partial_\n^\alpha\lambda 
\end{equation}
Next, we differentiate the above equation  with respect to $\partial_\n^\gamma$   and use (\ref{ee8}) together with  (\ref{es6}) to obtain
\begin{equation}
\label{ee10}
\partial_\n^\gamma\lambda=0
\end{equation}
Differentiating eq. (\ref{es2}) with respect to $\partial _\n^\gamma$ and combining  equations obtained by cyclic permutations of $\alpha,\beta,\gamma$ we arrive at 
\begin{equation}
\label{ee11}
\partial_\n^\gamma\partial_\n^\alpha a_\beta^\n=0
\end{equation}
and consequently 
\begin{equation}
\label{ee12}
a_\beta^\n=\frac{\lambda}{2}q_\beta^\n+U^\alpha_\beta q_\alpha^\n+d_\beta
\end{equation}
where $U=-U^T$ and $d_\beta$ do not depend on $q_\alpha^\n$.
\par 
Next, we will show inductively that for each $j=1,\ldots,\n$ we have
\begin{subequations}
\label{ee13}
\begin{equation}
\label{ee13a}
\quad \partial_j^\beta a^j_\alpha=\left(\frac{N+2}{2}-j\right) \lambda\delta^\beta_\alpha+U^\beta_\alpha,\\
\end{equation}
\begin{equation}
\label{ee13b}
\partial_j^\beta a=0,\quad \textrm{ and if } j>1 \textrm{ then }  \partial_j^\beta a^k_\alpha=0 \textrm{ for } k=1,\ldots,j-1
\end{equation}
\end{subequations}
Indeed, due to (\ref{es5}),(\ref{es6}) and (\ref{ee12})  for $j=\n$  the above assertion holds. Assume that it is true for  fixed $j$.  Then, 
by virtue of (\ref{es1})  and the induction hypothesis, for index $j$ we have
\begin{equation}
\label{ee14}
\partial_j^\alpha\partial_j^\beta\lambda=0
\end{equation}
On the other hand, differentiating (\ref{es7}) for $j-1$ with respect to $\partial_j^\gamma$   and using  the induction hypothesis we arrive at  
\begin{equation}
\label{ee15}
\partial_{j-1}^\gamma a^{j-1}_\alpha=(\frac{N+4}{2}-j)\lambda\delta^\gamma_\alpha+U^\gamma_\alpha+q^j_\alpha\partial_j^\gamma\lambda 
\end{equation}
Putting $\gamma=\alpha$ in  (\ref{ee15}) and  differentiating    with respect to $\partial_j^\beta$, one gets by (\ref{ee14}) and (\ref{ee13b})
\begin{equation}
\label{ee16}
\partial_j^\beta\lambda=0
\end{equation}
As a result  eq. (\ref{ee15}) takes the  form  of  (\ref{ee13a}) for $j-1$ and, due to (\ref{es1}), $\partial_{j-1}^\beta a=0$.  Moreover, differentiating  (\ref{es7}) for $k=1,\ldots,j-2$ with respect to $\partial_j^\beta$, by the induction hypothesis and (\ref{ee16}) one obtains $\partial ^\beta_{j-1} a_\alpha^k=0$ for $k=1,\ldots,j-2$  which completes the proof of (\ref{ee13}).
\par 
Additionally, by virtue of   (\ref{ee13b}) and (\ref{es1}) we conclude that $a$ and $\lambda$ are  functions of  $t$ only.
Differentiating eq. (\ref{ee13a}) with respect to $\partial_j^\beta$ (without summation)   and combining with equations obtained by cyclic permutations of $\alpha,\beta,\gamma$ we get $U_\beta^\alpha$ is also only  function of $t$.
\par
Now, differentiating (\ref{es7}) for $j=1,\ldots,\frac{N-1}{2}$ with respect to $\partial_j^\gamma$ (without summation) and using (\ref{ee13a})  we obtain a recurrence formula for $\partial_j^\gamma a_{j+1}^\alpha$
\begin{equation}
\label{ee18}
\partial_t\partial_j^\gamma a_\alpha^j+\underbrace{\partial_{j-1}^\gamma a^j_\alpha}_{0 \textrm{ for } j=1}=\partial_j^\gamma a^{j+1}_\alpha
\end{equation}
which, due to (\ref{ee13a}), can be explicitly solved   and the final result reads
\begin{equation}
\label{ee19}
 \partial_j^\gamma a_{j+1}^\alpha =(-j+N+1)\frac{j\dot\lambda}{2}\delta^\gamma_\alpha+j\dot U^\gamma_\alpha,\quad j=1,\ldots,\frac{N-1}{2}
\end{equation}
Similarly,  differentiating (\ref{es7}) with respect to $\partial_{j-1}^\gamma$ we obtain a recurrence formula  for $\partial _{j-1}^\gamma a_\alpha^{j+1}$, $j=2,\ldots,\frac{N-1}{2}$ which solution is of the form
\begin{equation}
\label{ee20}
\partial _{j-1}^\gamma a_\alpha^{j+1}=(-2j^2+3j(N+2)-3N-4)\frac{j\overset{..}{\lambda}}{12}\delta^\gamma_\alpha+\frac{(j-1)j}{2}\overset{..}{U}^\gamma_\alpha, \, 
\end{equation}
Let us now study  the behaviour of $c$. Differentiating (\ref{es4}) with respect to $\partial_\n^\delta\partial_\n^\gamma$ and using (\ref{ee19}) together with (\ref{ee13a})   we find that the third order derivative of $c$ with respect to  $q_\alpha^\n$ is zero, so 
\begin{equation}
\label{ee21}
c=c_1^{\alpha\beta} q_\alpha^\n q_\beta^\n+c_2^\alpha q_\alpha^\n+c_3
\end{equation}
where $c_1^{\alpha\beta}=c_1^{\beta\alpha},c_2^\alpha,c^3$ do not depend on $q_\alpha^\n$.
Substituting, (\ref{ee21}) into (\ref{es3}) and comparing terms with $q^\n$'s one gets the following set of equations
\begin{subequations}
\label{ee22}
\begin{align}
\label{ee22a}
0&=\partial^{(\gamma}_{\frac{N-1}{2}}c^1_{\alpha\beta)}\\
\label{ee22b}
0&=q^{k'+1}_\gamma\partial_{k'}^\gamma c_1^{\alpha\beta}+\partial^{(\beta}_{\frac{N-1}{2}}c_2^{\alpha)}+\partial_tc_1^{\alpha\beta}\\
\label{ee22c}
0&=q^{k'+1}_\gamma\partial_{k'}^\gamma c_2^{\alpha}+\partial^\alpha_{\frac{N-1}{2}}c^3+\partial_t c_2^\alpha\\
\label{ee22d}
0&=q^{k'+1}_\gamma\partial_{k'}^\gamma c_3+i\partial_t c_3+\frac{1}{m}c_1^{\alpha\alpha}
\end{align}
\end{subequations}
where index with  subscript " ' " runs from $1,\ldots,\frac{N-3}{2}$ and $(\alpha,\ldots)$ is  symmetrization over the  enclosed indices.
On the other hand, differentiating (\ref{es4}) with respect to  $\partial^\gamma_\n$  and next  substituting (\ref{ee12}) for $j=\n$ and (\ref{ee19}) for $j=\frac{N-1}{2}$  one gets
\begin{subequations}
\label{ee23}
\begin{equation}
\label{ee23a}
c_1^{\alpha\beta}=-\frac{mi}{16}(N+1)^2\dot\lambda\delta^{\gamma\alpha}
\end{equation}
\begin{equation}
\label{ee23b}
\dot U^\beta_\alpha=0
\end{equation}
\end{subequations}
Consequently $c_1$'s are functions of   $t$ only, using this fact, (\ref{ee19}),(\ref{es4})  and (\ref{ee20}) for $j=\frac{N-1}{2}$  we obtain
\begin{equation}
\label{ee24}
\partial^\gamma_{\frac{N-1}{2}}c^2_\alpha=-\frac{im}{48}\overset{..}{\lambda}\delta^\gamma_\alpha(N^2-1)(2N+3)
\end{equation}
Substituting (\ref{ee24}) and (\ref{ee23}) into (\ref{ee22b}), one finds that 
\begin{equation}
\label{ee25}
\overset{..}{\lambda}=0
\end{equation}
Thus, by virtue of (\ref{ee20}), we have
\begin{equation}
\label{ee26}
\partial _{j-1}^\gamma a_\alpha^{j+1}=0,  \quad j=2,\ldots,\frac{N-1}{2}   
\end{equation}
A simple  consequence of (\ref{ee26}) is 
\begin{equation}
\label{ee26b}
\partial_1^\gamma a_\alpha^j=0, \quad j=3,\ldots,\n
\end{equation}
 Indeed, differentiating (\ref{es7}) with respect to  $\partial_1^\gamma$ 
we have $(\partial_t+q_\beta^{k+1}\partial_k^\beta)(\partial_1^\gamma a_\alpha^j)=\partial_1^\gamma a_\alpha^{j+1} $. This together with eq. (\ref{ee26}) for $j=2$ imply (\ref{ee26b}).
\par
Now, we show that for each $k=2,\ldots,\frac{N-1}{2}$ we have
\begin{equation}
\label{ee27}
\partial_{j-k}^\gamma a^j_\alpha=0,\quad j=k+1,\ldots,\n
\end{equation}
Indeed, for $k=2$ eq.  (\ref{ee27}) holds due to (\ref{ee26}). Assume now  (\ref{ee27}) is true for $k-1$; we will show that  it holds for $k$.
Differentiating (\ref{es7}) with respect to $\partial^\gamma_{j-k+1}$,  by the induction hypothesis, one obtains
\begin{equation}
\label{ee28}
\partial_{j-k}^\gamma a_\alpha^j=\partial^\gamma_{j-k+1} a_\alpha^{j+1},\quad j=k+1,\ldots,\n
\end{equation}
but for $j=k+1$, due to (\ref{ee26b}),  we have $\partial_{1}^\gamma a_\alpha^{k+1}=0$ which proves (\ref{ee27}).
\par 
Let us note that (\ref{ee27}) implies $\partial_k^\gamma a_\alpha^\n=0$ for $k=1,\ldots,\frac{N-3}{2}$; therefore 
\begin{equation}
\label{ee29}
a_\alpha^\n=\frac{\lambda}{2}q^\n_\alpha+\dot\lambda\frac{(N+3)(N-1)}{8}q_\alpha^{\frac{N-1}{2}}+U_\alpha^\beta q^\n_\beta+f_\alpha(t)
\end{equation}
Substituting this in (\ref{es4}) we find that $c^2_\alpha$ is  a function of $t$ only, more precisely we have
\begin{equation}
\label{ee30}
c^2_\alpha=-im \dot f_\alpha
\end{equation}
Furthermore, the partial derivatives of $c_3$ are expressed in terms of $c_2$:
\begin{equation}
\label{ee31}
\partial_k^\alpha c^3=(-1)^\frac{N-2k+1}{2}\partial_t^{(\frac{N-2k+1}{2})}c^2_\alpha,\quad k=1,\ldots,\frac{N-1}{2}
\end{equation}
this can be proved inductively  differentiating eq. (\ref{ee22d}) with respect to $\partial^\alpha_{\frac{N-1}{2}-l}$ for   $l=0,\ldots,\frac{N-3}{2}$ and using eqs. (\ref{ee22c}), (\ref{ee30}).
Differentiating (\ref{ee22d}) with respect to  $\partial_1^\gamma$ and using  eq. (\ref{ee31}) one gets
\begin{equation}
\label{ee32}
\partial_t^{(\n)}c^2_\alpha=0
\end{equation}
Thus, by (\ref{ee30}), we have
\begin{equation}
\label{ee32b}
c^2_\alpha=\sum_{k=0}^{\frac{N-1}{2}}A_\alpha^k t^k, \quad f_\alpha=\frac{i}{m}\sum_{k=0}^{\frac{N-1}{2}}A^k_\alpha \frac{t^{k+1}}{k+1}+D
\end{equation}
where $A_\alpha^k$ and $D$ are some constants.
\par
Summarizing, due to (\ref{ee22d})  and (\ref{ee23a}), (\ref{ee31}) we have 
\begin{equation}
\label{ee33}
c_3=\frac{(N+1)^2}{16}d\lambda-(-1)^{\frac{N-2k'+1}{2}}q_\alpha^{k'+1}\partial_0^{(\frac{N-2k'-1}{2})}c^2_\alpha+C
\end{equation}
where $C$ is a constant.
\par
Now, it remains to find the dependence $a_\alpha^j$ of $t$.  By virtue of (\ref{es7}) and equations (\ref{ee13}),(\ref{ee19}) and (\ref{ee26})
we  have
\begin{equation}
\label{ee34}
a_\alpha^{j+1}=\partial_t a_\alpha^j-f_\alpha^j, \quad j=1,\ldots,\frac{N-1}{2}
\end{equation}
where $f^j_\alpha=\lambda(j-\frac{N}{2})q^{j+1}_\alpha+\frac{\dot\lambda}{2}(j-N-2)(j-1)q_\alpha^j-U_\alpha^\beta q_\beta^{j+1}$ and by definition $f^j_\alpha=0$ for $j<1$.
Since $\overset{..}{\lambda}=0$,  the explicit solution of (\ref{ee34}) is of the form 
\begin{equation}
\label{ee35}
a_\alpha^{j}=\partial_t^{(j-1)}a^1_\alpha -\partial_tf_\alpha^{j-2}-f_\alpha^{j-1}, \quad j=2,\ldots,\frac{N+1}{2}
\end{equation}
Comparing eq. (\ref{ee35}) for $j=\n$ with eq. (\ref{ee29}) we obtain that $f_\alpha=\partial_t^{(\frac{N-1}{2})}a_\alpha^1$  and consequently, by virtue of (\ref{ee13}) and (\ref{ee25})
\begin{equation}
\label{ee36}
a^1_\alpha=\sum_{k=0}^{\frac{N-3}{2}}B_\alpha^k t^k+\frac{N}{2}\lambda q^1_\alpha +U^\beta_\alpha q_\beta^1+\underbrace{\int\ldots\int }_{\frac {N-1}{2}}f_\alpha dt
\end{equation}
where $G_\alpha^k$ are some constants. Coming  back to  (\ref{ee35})   we arrive at 
\begin{equation}
\label{ee37}
\begin{split}
&a_\alpha^j=\sum_{l=0}^{\frac{N-1}{2}-j}B_\alpha^{j+l-1}\frac{(j+l-1)!}{l!}t^l+\frac{i}{m}\sum_{k=0}^{\frac{N-1}{2}}A^k_\alpha \frac{k!t^{k-j+\frac{N+3}{2}}}{(k-j+\frac{N+3}{2})!}+\sum_{l=0}^{\frac{N-1}{2}-j}C^lt^l\\
&+D\frac{t^{\frac{N+1}{2}-j}}{\frac{N+1}{2}-j}+\frac{\dot\lambda}{2}(j-1)(N-j+2)q_\alpha^{j-1}+\frac{\lambda}{2}(N-2j+2)q_\alpha^j+U^\beta_\alpha q_\beta^j
\end{split} 
\end{equation}
Shifting  $B$'s by $C$'s or $D$,  after some indices manipulations we find that $a_\alpha^j$, for $j=1,\ldots,n$ take the form
\begin{equation}
\label{ee38}
\begin{split}
&a_\alpha^j=\sum_{l=j-1}^{\frac{N-1}{2}}B_\alpha^{l}\frac{l!}{(l-j+1)}t^{l-j+1}+\frac{i}{m}\sum_{k=0}^{\frac{N-1}{2}}A^k_\alpha \frac{k!t^{k-j+\frac{N+3}{2}}}{(k-j+\frac{N+3}{2})!}\\
&+\frac{\dot\lambda}{2}(j-1)(N-j+2)q_\alpha^{j-1}+\frac{\lambda}{2}(N-2j+2)q_\alpha^j+U^\beta_\alpha q_\beta^j
\end{split}
\end{equation}
Due to eq.(\ref{ee25})  $\lambda=2Et+F$ thus  (\ref{es1}) yields  $a=Et^2+Ft+G$. 
\par
Summarizing, we see  that all $a$'s and $c$ depend on some constants $C,E,F,G$, $A$'s, $B$'s and $U$'s. Thus the maximal symmetry  algebra  of
(\ref{ee3}) is finite-dimensional and  its basis is obtained by selecting the coefficient related to  these constants. After troublesome indices manipulations, we find all generators:
\begin{subequations}
\par  $iG$
\begin{equation}
\label{ee39a}
H=-i\partial_t
\end{equation}
\par  $iF$
\begin{equation}
\label{ee39b}
D=-it\partial_t-i\sum_{j=1}^\n(\frac{N}{2}-j+1)q_\alpha^j\partial_j^\alpha-i\frac{1}{16}(N+1)^2d
\end{equation}
\par $iE$
\begin{equation}
\label{ee39c}
\begin{split}
&K=-it^2\partial_t-i\frac{1}{8}(N+1)^2dt-i\sum_{i=1}^\n\left((j-1)(N-j+2)q_\alpha^{j-1}\partial_j\right.\\
&\left. -it(N-2j+2)q_\alpha^j\partial_j^\alpha\right)-m\frac{(N+1)^2}{8}q^{\n}_\alpha q^{\n}_\alpha
\end{split}
\end{equation}
\par $iB_\alpha^l$, for  $l=0,\ldots,\frac{N-1}{2}$
\begin{equation}
\label{ee39d}
 P_l^\alpha=-il!(\sum_{k=0}^l\frac{t^{l-k}}{(l-k)!}\partial_{k+1}^\alpha)\\
\end{equation}
\par $-\frac{(j-\n)!}{mj!}A^\alpha_{j-\n}$, for $j=\n,\ldots,N$
\begin{equation}
\label{ee39e}
P_j^\alpha=-ij!\sum_{l=0}^{\frac{N-1}{2}}\frac{t^{j-l}}{(j-l)!}\partial_{l+1}^\alpha-mj!\sum_{k=\n}^j q_\alpha^{N-k+1}(-1)^{\n+k}\frac{t^{j-k}}{(j-k)!},\\
\end{equation}
\par
$\frac{i}{2}U_\alpha^\beta$
\begin{equation}
\label{ee39f}
J_\beta^\alpha=-i(q_\beta^j\partial_j^\alpha-q_\alpha^j\partial_j^\beta)
\end{equation}
$\frac{C}{m}$
\begin{equation}
\label{ee39g}
Z=m
\end{equation}
\end{subequations}
The obtained generators agree with the ones  from \cite{b6} and do satisfy $N$-GCA commutation rules with central charge $Z=m$. Thus  centrally extended $N$-GCA is in fact the maximal symmetry algebra of  the Sch\"odinger equation corresponding to the free theory with higher order time derivatives. 
\par
{\bf Acknowledgments} The discussions with Cezary Gonera and
Pawe\l \  Ma\'slanka are gratefully acknowledged. This work  is supported  in part by  MNiSzW grant No. N202331139.

\end{document}